\pdfoutput=1

\documentclass[11pt]{article}

\usepackage[]{acl}

\usepackage{times}
\usepackage{latexsym}
\usepackage{booktabs}
\usepackage{float}

\usepackage[T1]{fontenc}

\usepackage[utf8]{inputenc}

\usepackage{microtype}

\usepackage{inconsolata}

\usepackage{graphicx}
\usepackage{enumitem}
\usepackage{xcolor}

\title{What are the limits of cross-lingual dense passage retrieval for low-resource languages?}

\author{Jie Wu \\
  Leiden University\\
  \texttt{} \\\And
  Zhaochun Ren \\
  Leiden University\\
  \texttt{\small{z.ren@liacs.leidenuniv.nl}} \\\And
  Suzan Verberne \\
    Leiden University\\
  \texttt{\small{s.verberne@liacs.leidenuniv.nl}}
  \\}

\begin{document}
\maketitle
\begin{abstract}
In this paper, we analyze the capabilities of the multi-lingual Dense Passage Retriever (mDPR) for extremely low-resource languages.  
In the Cross-lingual Open-Retrieval Answer Generation (CORA) pipeline, mDPR achieves success on multilingual open QA benchmarks across 26 languages, of which 9 were unseen during training. These results are promising for Question Answering (QA) for low-resource languages. 
We focus on two extremely low-resource languages for which mDPR performs poorly: Amharic and Khmer. We collect and curate datasets to train mDPR models using Translation Language Modeling (TLM) and question--passage alignment. We also investigate the effect of our extension on the language distribution in the retrieval results.
Our results on the MKQA and AmQA datasets show that language alignment brings improvements to mDPR for the low-resource languages, but the improvements are modest and the results remain low. 
We conclude that fulfilling CORA's promise to enable multilingual open QA in extremely low-resource settings is challenging because the model, the data, and the evaluation approach are intertwined. Hence, all three need attention in follow-up work. 
We release our code for reproducibility and future work.\footnote{Anonymized git repository for peer review: \url{https://anonymous.4open.science/r/Question-Answering-for-Low-Resource-Languages-B13C/}} 
\end{abstract}

\section{Introduction}
Due to the limited training data for low-resource languages, the effectiveness of Question Answering (QA) models for those languages is typically low. 
One of the most successful retrievers for QA is the Dense Passage Retriever (DPR)~\cite{karpukhin2020dense}, a model that is trained on English question and answer passage pairs using a pre-trained BERT model with a dual-encoder architecture. After the English-only DPR, a multilingual Dense Passage Retriever (mDPR) was introduced by~\citet{asai2021one} for multilingual and cross-lingual settings. 
mDPR uses multilingual BERT (mBERT) instead of the English-only BERT, thereby extending DPR to a multilingual setting. 

In this paper, we 
aim to improve the quality of mDPR for low-resource languages that are not included in the pre-training of mBERT, specifically Amharic and Khmer. To this end, we construct training data sets using machine translation and extend the vocabulary of the models. We then post-train the retrieval model with Translation Language Modeling (TLM)~\cite{conneau2019cross}, and fine-tune it with cross-lingual question--passage alignment. TLM is a self-supervised learning method that can be used to map sentences of low-resource languages to sentences of high-resource languages (in our case Thai, Arabic, and English). We investigate the effect of cross-lingual training on Amharic and Khmer.

We evaluate three models with different language alignments: with a related high-resource language (Amharic--Arabic and Khmer--Thai); with a non-related high-resource language (Amharic--Thai and Khmer--Arabic); and with English. We also analyze the effect of language alignment on the cross-lingual retrieval results: what are the languages represented in the results when the ranker receives a query in Amharic or Khmer?

Our contributions are as follows:
\begin{itemize}[noitemsep,nolistsep]
\item We extend mDPR by additional post-training for two low-resource languages, Amharic and Khmer. 
    \item We provide curated datasets containing aligned sentences between 6 language pairs from the CCAligned dataset. 
    \item We provide a dataset with aligned questions and passages for 6 language pairs.
    \item We show that despite our language alignment methods can improve QA quality for low-resource languages, the effectiveness of mDPR for low-resource languages remains limited.
\end{itemize}
To facilitate the reproducibility and follow-up of the results reported in this paper, we have released our code.$^1$

\section{Related Work}

\textbf{QA for low-resource languages}
In order to improve QA models for low-resource languages, data augmentation techniques have been proposed~\cite{kumar2022mucot, pramana2022improving, singh2019xlda, riabi2021synthetic}. For example, to improve the QA performance in two low-resource languages Hindi and Tamil, \citet{kumar2022mucot} propose a training pipeline with  
data augmentation based on translations and transliterations. 
They found that the translation of the same language family increases the model performance, whereas transliteration does not increase the performance. 

Other studies on low-resource QA include 1) the creation of datasets for low-resource languages such as Swahili \cite{wanjawa2023kenswquad}, Telugu \cite{vemula2022tequad}, Persian \cite{darvishi2023pquad} Vietnamese \cite{le2022vimqa,vuong2023improving}, and a comprehensive effort for 15 African languages \cite{ogundepo2022africlirmatrix}; 2) QA systems that are developed for a specific low-resource language, such as for Bengali \cite{das2022question, banerjee2014bfqa}, Vietnamese \cite{phan2021building, do2021developing}, Telugu \cite{ravva2020avadhan}, Persian \cite{mozafari2022peransel} and Tamil \cite{antony2023question}; and 3) improving QA systems for low-resource languages with, e.g., unsupervised statistical methods \cite{das2022improvement}, syntactic graphs \cite{wu2022enhancing}, and cross-lingual token alignment \cite{huang2023improving}.

\citet{bonab2019simulating} analyze the factors that influence cross-lingual information retrieval (CLIR) quality. They find that language families are relevant and that for both the low-resource and the high-resource language, good quality training data are required. In order to create better representations in CLIR, \citet{fei2021cross} propose to add a cross-\textit{modal} dimension, including images in the parallel training. \citet{yarmohammadi2019robust} have achieved improved document representations for spoken documents, specifically for spoken data in Somali, Swahili, and Tagalog to be retrieved by English queries. \citet{wang2020extending} propose to the vocabulary of mBERT to low-resource languages and further post-train the model with the extended vocabulary on the corresponding low-resource languages. They obtain 
performance boosts of mBERT on the NER task for the corresponding low-resource languages.

Unlike previous studies, we aim to improve the retrieval performance of two low-resource languages (i.e., Amharic and Khmer) through translation-based post-training of the mBERT model: we use the TLM paradigm to map sentences of those low-resource languages to their corresponding sentences in high-resource languages. The post-trained mBERT model is then fine-tuned through question--passage alignment between the language pairs and evaluated on the QA datasets.

\textbf{Post-training}
Post-training or post-pretraining is the step between the initial pre-training and fine-tuning, which can be applied when pre-training the model from scratch is too expensive. \citet{yu2021cross} create distant supervision data from multilingual Wikipedia pages through section alignment. \citet{pan2021multilingual} propose a method called PPA, which stands for Post-Pretraining Alignment where self-supervised word- and sentence-level alignment are applied as a post-pretraining step to improve cross-lingual transferability of the mBERT model using parallel data. In their method, Translation Language Modeling (TLM) \cite{conneau2019cross} is used to align multilingual contextual embeddings on the word level, and contrastive learning and random shuffling are used to align embeddings on the sentence level. Their results show that PPA efficiently improves the cross-lingual transferability of Language Models. Also, the proposed method shows to largely improve the performance of mBERT on Natural Language Inference (NLI) and QA tasks \cite{pan2021multilingual}.


\section{Data collection and curation}

\subsection{Sentence alignment data}
We use the CCAligned dataset\footnote{\url{https://www.statmt.org/cc-aligned/}} \cite{el2020ccaligned} to create training data for the sentence-alignment task (TLM). 
We use the sentence pairs in the CCAligned dataset for Arabic--English, Amharic--English, Thai--English, and Khmer--English. 
We create non-English sentence pairs by joining them on the English sentences to obtain candidate Amharic--Arabic and Khmer--Thai sentence pairs, as well as candidate Amharic--Thai and Khmer--Arabic sentence pairs that we filtered as described next.



\textbf{Creating the Amharic datasets} 
The original English-Arabic dataset contains 25,309,750 sentence pairs, while the original English-Amharic dataset contains 346,517 sentence pairs, and by joining them, we obtain 8,817,926 sentence pairs for Amharic and Arabic. However, we found that some sentences in Arabic are not aligned with the corresponding sentences in Amharic. In order to filter out the noise, we translate the original Amharic sentences into English 
with the DL Translate tool\footnote{\url{https://github.com/xhluca/dl-translate}} 
and then filter the translations based on the sentence similarity\footnote{\url{https://github.com/Susheel-1999/Sentence_Similarity}} using the \texttt{all-MiniLM-L6-v2} model 
between the translated English sentences (from Amharic) and the original English sentences. 
Based on a manually assessed random sample of 50 pairs, 
we set the threshold on the cosine similarity at 0.7 to select the sentence pairs that are sufficiently similar to each other. 
The filtered Amharic--English dataset contains 149,573 Amharic--English sentence pairs. 

Using the same strategy on Arabic--English sentences, we obtain a threshold of 0.8 and a filtered Arabic--English dataset containing 87,970 Arabic--English sentence pairs that have an Amharic counterpart. 
Finally, we merge the filtered Amharic--English dataset with the filtered Arabic--English dataset and obtain a joined dataset with English, Arabic, and Amharic sentences, containing 
102,994 unique sentence pairs for the Amharic--Arabic 
sentence alignment. For the Amharic--English sentence alignment, we use the filtered Amharic--English dataset containing 149,573 sentence pairs. However, to make the data size comparable to the Amharic--Arabic dataset, we only selected the Amharic--English sentence pairs with a similarity score equal to or higher than 0.8 to form our Amharic--English sentence alignment dataset, which gives us 99,403 sentence pairs.

\textbf{Creating the Khmer datasets}
The same process is applied to Khmer and Thai. The original English-Khmer dataset contains 412,381 sentence pairs and the English-Thai dataset contains 10,746,367 sentence pairs. The join on English sentences results in 5,510,283 sentence pairs. 
Using the same filtering method as above, we get a similarity threshold of 0.8 for Khmer and obtain a filtered Khmer--English dataset containing 119,757 sentence pairs. 

Similarly to Arabic--English, we obtain 114,394 Thai--English sentence pairs that have a Khmer counterpart.  
After translation and with a similarity threshold of 0.8, we obtain a filtered Thai--English dataset containing 68,833 sentence pairs. Finally, 
we obtain a joined dataset with English, Thai, and Khmer sentences, containing 
80,935 unique instances which are used for Khmer--Thai sentence alignment. We use the filtered Khmer--English dataset containing 119,757 instances for Khmer--English sentence alignment.

\textbf{Cross-paired datasets}
The same process is repeated to create the Amharic--Thai and Khmer--Arabic datasets for comparison. For Amharic--Thai, the resulting dataset has 77,413 instances and for Khmer--Arabic, we have 106,148 instances. 

\subsection{Data for mDPR fine-tuning}\label{mDPR-data}
\textbf{Training and development data}
For mDPR fine-tuning, we need training data including the three high-resource languages: English, Arabic, and Thai. Therefore, we created an extended version of the MIA training data \cite{asai2022mia}, including Thai QA data as well. We obtained the Thai data from the XQuAD\footnote{\url{https://github.com/google-deepmind/xquad}} dataset \cite{artetxe2020cross} and the iapp-wiki-qa-squad\footnote{\url{https://github.com/iapp-technology/iapp-wiki-qa-dataset}} dataset \cite{Viriyayudhakorn2021iapp}. 
We combine these two datasets into a Thai dataset with 8,432 examples. Detailed statistics are shown in Table~\ref{tab:thai_data} in the Appendix. 
Each training example consists (at least) of a question, the answers, the positive contexts, a question ID, and the language. Each validation example has similar entries as the training data, only the language is left out. 

\textbf{Question--passage alignment for low-resource languages} Since we have limited QA data in Amharic and Khmer, we take 5,000 training examples from the above-mentioned training data in the corresponding high-resource language. We translate the questions in the training samples into the corresponding low-resource language using Google Translate.\footnote{\url{https://pypi.org/project/deep-translator/}} 
We add these training examples to the above-mentioned training data. 

\textbf{Evaluation data}
\citet{asai2022mia} provide the XOR-TyDi development data and MKQA development data. 
We also add the above mentioned Thai development data to this XOR-TyDi development data. 
Since the original MKQA dataset provided by \citet{longpre2021mkqa} also includes Thai, we add the corresponding Thai examples to the MIA version of the MKQA dataset. 

We evaluate the model performance for Amharic and Khmer on two QA datasets for these languages, AmQA\footnote{\url{https://github.com/semantic-systems/amharic-qa}} \cite{abedissa2023amqa} and MKQA\footnote{\url{https://github.com/apple/ml-mkqa/tree/main}} \cite{longpre2021mkqa}. 
The Khmer test data are taken from the original MKQA dataset (leaving out the used development data), and consist of 5,000 examples. The Amharic test data are taken from the Amharic Question Answering Dataset (AmQA) and consist of 2,622 examples. Each test data example has the same entries as the development data (used for evaluation) described above.

\section{Methods}

\subsection{mDPR}
Our research builds on the CORA model proposed by~\citet{asai2021one}. The CORA model consists of two parts: 1) Multilingual Dense Passage Retriever (mDPR), and 2) Multilingual Answer Generator (mGEN). 
The mDPR part of the CORA model is based on the DPR (Dense Passage Retrieval) model proposed by~\citet{karpukhin2020dense}. 
mDPR employs multilingual BERT to represent passages and questions.  
The initial training of mDPR was based on two QA datasets: 1) Natural Questions (NQ) \cite{kwiatkowski2019natural}, and 2) XOR-TyDi QA and TyDi QA \cite{clark2020tydi}. 



\subsection{Language alignment intuitions}

The idea of post-training with language alignment is that by encoding text pairs of two languages, the model can learn the interactions between the words in these texts~\cite{conneau2019cross}. 
The model can then better map the question to the documents of the other language  and thereby improve the cross-lingual retrieval effectiveness, especially for low-resource languages. 


We compare three alignment pairs: (1) alignment between a low-resource and a related high-resource languages (Amharic with Arabic and Khmer with Thai); (2) alignment between a low-resource and a non-related/more distant high-resource languages (Amharic with Thai and Khmer with Arabic); (3) alignment between a low-resource language (Amharic or Khmer) and English. The first variant is our proposed method for improving mDPR for low-resource languages. The second and third variants are used as experimental comparisons to evaluate the effectiveness of our proposed method.

The reason that we choose Amharic and Khmer as low-resource languages for our research is that 1) sentence-alignment data is available in the CCAligned dataset \cite{el2020ccaligned}, 2) QA evaluation data is available, and 3) both languages are not included in mBERT. 
We choose Thai as the related high-resource language for Khmer because a) the shared word vocabulary between the two languages is about 30\%, b) their syntax is almost the same, and c) many idioms are shared between these two languages \cite{Michael2023Thai}. The choice of Arabic as the related high-resource language for Amharic is because 1) we cannot find any other high-resource languages that are closer to Amharic than Arabic, and 2) they come from the same Afroasiatic language family (the Semitic branch), which means that there is some similarity between the two languages. However, they do have a different writing system and script. Both high-resource languages, Thai and Arabic, are included in mBERT-cased (\texttt{bert-base-multilingual-cased}), which we use as our retrieval model for mDPR. The original retrieval model used by CORA is mBERT-uncased, which does not include Thai in the pre-training of the model. Since there is no uppercase and lowercase in the languages Khmer and Amharic (same for Thai and Arabic), we do not expect that the cased version of mBERT will have a substantial effect on the performance of these languages.

\subsection{mBERT Post-training}
The purpose of post-training mBERT is to allow the model to create representations for texts in Amharic and Khmer, which were both not included in the pre-training data of mBERT. Our post-training step consists of two learning objectives: Masked Language Modeling (MLM) and Translation Language Modeling (TLM).

Since mBERT is not trained in Amharic and Khmer, the corresponding vocabulary also does not contain Amharic and Khmer tokens as both scripts are different from those in the original mBERT model. 
Consequently, due to a large number of `[UNK]' tokens with all having the same embedding~\cite{toraman2023impact}, the learning might not be effective. 
Our proposed solution involves adding Amharic and Khmer tokens to the mBERT vocabulary, allowing the embeddings for these newly added tokens to be learned through the MLM and TLM tasks. By doing so, we create two new mBERT models, \texttt{amBERT} and \texttt{kmBERT}, as follows:

\textbf{1. Extending the vocabulary.} For \textbf{Amharic}, we first get all Amharic sentences from the Amharic--Arabic and Amharic--Thai datasets. Since there is already a trained mBERT model that is fine-tuned on Amharic texts (where the original mBERT vocabulary is replaced by an Amharic vocabulary)\footnote{\url{https://huggingface.co/Davlan/bert-base-multilingual-cased-finetuned-amharic}}, we use the tokenizer of this Amharic-BERT model to tokenize all Amharic sentences. From the obtained list of Amharic tokens, we only keep those full Amharic words that are unknown by the original mBERT model. This is done by retokenizing each Amharic token using the original mBERT model. 
By doing so, we obtain a total of 37,852 unique Amharic tokens. 
We will refer to our extended mBERT model with Amharic as the amBERT model.

For \textbf{Khmer}, similarly, we get all Khmer sentences from the Khmer--Thai and Khmer--Arabic datasets. 
Since Khmer tokens are not separated by whitespace,  we have to preprocess the Khmer sentences before applying our extended mBERT tokenizer such that the sentences can be successfully encoded. We use the word tokenizer from \texttt{khmer-nltk}\footnote{\url{https://pypi.org/project/khmer-nltk/\#description}} to tokenize all Khmer sentences. 
We then retokenize the Khmer tokens with the original mBERT tokenizer 
and only add tokens to our mBERT vocabulary that are unknown by the original mBERT model. This way, we obtain 22,164 unique Khmer tokens. When our mBERT vocabulary is extended with these Khmer tokens, we can use the corresponding tokenizer to tokenize Khmer sentences that are already separated by whitespace. 
We will refer to our extended mBERT model with Khmer as the kmBERT model. See Table~\ref{tab:tokenization} in the appendix for tokenization examples.

After we have obtained our \texttt{amBERT} tokenizer and \texttt{kmBERT} tokenizer, the embedding size of the corresponding mBERT model is resized such that it is equal to the length of the corresponding vocabulary size. The embedding of these newly added tokens will be randomly initialized and learned through the post-training process.

\textbf{2. MLM. } For Masked Language Modeling (MLM) we use the documents of the low-resource languages (i.e., Amharic and Khmer) obtained from the CCAligned dataset~\cite{el2020ccaligned} as input data. The statistics are in Table~\ref{tab:doc_distribution} in the Appendix. 
The MLM task is used to learn the representations of the newly added Amharic/Khmer words. We randomly mask 15\% of the input data.  
Based on recommendations of the BERT authors \cite{devlin2019bert} we split the documents into chunks of 128 tokens and chunks of 512 tokens, separately, before applying the masking.\footnote{\url{https://github.com/google-research/bert/tree/master}}
We use a sequence length of 128 for the two low-resource languages, a training batch size of 32, evaluation batch size of 8, and Adam as our optimizer with a learning rate of $2 \times 10^{-5}$. A linear learning schedule with warmup is used where the learning rate linearly decreases from the initial learning rate to 0 after a warmup period. Here, we use a weight decay of 0.01 and warmup steps of 10,000. At the start of the post-training process, the model for each language pair is initialized with the weights of the original mBERT model and post-trained for 500,000 steps. For reproduction purposes, we set the random seed to 12345. Then, we additionally post-train the resulting mBERT model(s) with MLM using a sequence length of 512. For this, we use a training batch size of 8 and we trained the model for another 100,000 steps. 

\textbf{3. TLM.} We adopt Translation Language Modeling (TLM) \cite{conneau2019cross} to learn better representations of tokens in the low-resource languages by aligning sentences with sentences in high-resource languages. Specifically, we combine each sentence from the low-resource language with the corresponding sentence from the high-resource language to form one single sequence. Here, we have two variants of each sentence pair, one with first the sentence from the low-resource language, and one with first the sentence from the high-resource language. This gives us more data to train on and variation between the position of the low- and high-resource languages. We again mask 15\% of the input data. 
We truncate with a maximum length of 256. 
We encode the sentence pairs with the mBERT tokenizer into our mBERT model. 
We first preprocess the Khmer sentences to separate the words by whitespace using khmer-nltk. 
We use a training batch size of 16. The model(s) is trained for 300,000 steps. The other hyper-parameter values are the same as described in the previous paragraph.

After the post-training process, we obtain six models with extended vocabularies for the following language pairs: 1) Amharic--Arabic, 2) Amharic--Thai, 3) Amharic--English, 4) Khmer--Thai, 5) Khmer--Arabic, and 6) Khmer--English.

\subsection{mDPR Finetuning}

The post-trained mBERT model is further fine-tuned using question--passage alignment between the language pairs and evaluated on the QA datasets. So, we will have the questions in the low-resource language and passages in the high-resource language to improve the matching in their dense embeddings. In this way, we can improve the transferability between those low-resource languages such that only training data in high-resource languages (and limited training data in the low-resource languages) suffice for the QA task. 

\section{Experiments}


We use the hyperparameter values 
of \textit{replication 1} from MIA, based on the DPR hyperparameters \cite{karpukhin2020dense}. 
Due to limited GPU resources, we use a single GPU to train the models and apply a half-precision floating point format (FP16) instead of a single-precision floating point format (FP32) during training. 
After training, we use our trained mDPR models to generate dense embeddings for the Wikipedia context passages. 

More details on fine-tuning and data preparation are in Appendix section \ref{sec:details}.

\textbf{Evaluation.} For evaluating our mDPR models on the two low-resource languages Amharic and Khmer, we analyse if the answer(s) to the questions can be found in the top 20 passages based on the regex match. The string match is done by normalizing the context and answer(s) and then checking whether the sequence of answer tokens appears in the context. The regex match is done using the pattern search method to check whether the (normalized) answer(s) appears in the context. We then compute answer-level Recall@10 and answer-level Recall@20. In addition, as we assume that some passages might not contain the whole answer, but a part of the answer, we measure ROUGE-1. We compute the ROUGE-1 between the ground truth answer and each retrieved passage; then we take the highest value among all retrieved passages as the result for that question, after which we average over questions.

\textbf{Baseline.}
As our main baseline, we use the mDPR model based on the original mBERT-cased (\texttt{bert-base-multilingual-cased}) model without any modification or post-training (combined with the mGEN model provided by \citet{asai2022mia} in the MIA repository). The baseline model was fine-tuned and evaluated in the same way as described in the previous sections, but for the baseline, we do not apply any form of language alignment.

\section{Results}

\subsection{Main results}


\begin{table*}[t]
    \caption{Retrieval performance of different mDPR models for Amharic and Khmer in terms of Recall. * means that the result is significantly different ($P<0.05$) from the baseline model according to McNemar's paired test. 
    }
    \centering
    \begin{tabular}{|l||cccc|cccc|}
         \toprule
         &  \multicolumn{4}{|c|}{\textbf{Recall@10 (\%)}} & \multicolumn{4}{|c|}{\textbf{Recall@20 (\%)}}  \\
\toprule
\textbf{Amharic} &  & am-en & am-ar & am-th &  & am-en & am-ar & am-th\\
\hline
mBERT & 0.46 &  &  &  & 0.72 &  &  & \\
+ Posttraining &  & 2.10* & 2.10* & \textbf{2.25}* &  & 2.82* & \textbf{3.28}* & 3.09*\\
+ FT align  &  & 1.41* & 0.72 & 0.50 &  & 1.87* & 0.88 & 0.69\\
\midrule
\textbf{Khmer}  &  & km-en & km-ar & km-th &  & km-en & km-ar & km-th\\
\hline
mBERT & 7.74 &  &  &  & 9.94 &  &  & \\
+ Posttraining &  & 6.76* & 5.62* & 7.70 &  & 8.56* & 7.18* & 9.62\\
+ FT align  &  & \textbf{18.64*} & 8.30 & 9.50* &  & \textbf{22.00*} & 10.54 & 11.30*\\
\bottomrule
    \end{tabular}
    \label{am_results1}
\end{table*}

\textbf{Effect of post-training on Amharic} 
The results in Table \ref{am_results1} show that after posttraining using MLM and TLM, all three Amharic models trained using outperform the baseline. We only show recall results here, but the results in terms of ROUGE-1 provide the same pattern (Appendix Table \ref{rouge_results}). 
The Amharic--Thai model appears to be slightly better than the other two language pairs according to answer-level Recall@10, and the Amharic--Arabic model is slightly better according to answer-level Recall@20. The table also shows that the results after fine-tuning alignment are lower than with only the posttraining. This is probably because the fine-tuning alignment learns to match between the low- and high-resource languages. As a result, after fine-tuning, the model is more likely to retrieve passages in the corresponding high-resource languages (see Figure~\ref{fig:am_lang_distr} in the appendix). Thus, the finetuned model is disadvantaged when evaluated against Amharic ground truth. However we noticed that all results are low, so it remains difficult for the model to learn the embeddings in Amharic in order to perform well on this QA task.

\textbf{Effect of post-training on Khmer} For the Khmer models, we see in Table \ref{am_results1} that posttraining does not lead to better models than the baseline. (Again, the ROUGE-1 results are in the Appendix, Table \ref{rouge_results}.) For Khmer, the Khmer--English model with fine-tuning alignment is the best-performing model. 
After fine-tuning alignment, the best-performing model is the Khmer--English model. This is possibly due to the fact that the Khmer QA data is sampled from the Natural Questions dataset that was originally given in English. Further, we also noticed that Khmer results are better than those of Amharic, this might be due to the fact that the Khmer QA data includes multiple answers (aliases) for some questions, and these answers might also be given in the high-resource language such as English. Whereas for Amharic, only a single answer is given per question (mostly given in Amharic), and therefore more difficult to find the correct answer in the retrieved passages compared to Khmer.

\textbf{Comparing language alignment choices} For both low-resource languages, we find that the best-performing TLM model is the model with English, and the model with the related high-resource language is comparable to the model with the more distant high-resource language. Although the difference in performance is small for Amharic, we can say that for these two low-resource languages, the data size of the high-resource language is more useful than the relatedness between the language pair for improving the QA performance of the low-resource languages.

\textbf{Additional findings} 
In the appendix Table \ref{xor_res}, we show as additional results the effect of post-training on the higher-resource languages in the data. It appears that for almost all languages, the quality of our models is lower than the baseline. The difference in the overall F1 between our models and the baseline is around 3-5\% points. This indicates that language alignment between the language pairs brings some confusion to the models, lowering the effectiveness for the high-resource languages. 

\subsection{Analysis of cross-lingual retrieval}

We analyze the language distribution of the top-20 retrieved passages by the different models for questions in Amharic. 
For the baseline model, over 90\% of the documents that are retrieved, are Thai. This is surprising as we assume that Amharic is not similar to Thai. We speculate that this is caused by the mBERT model finding no passages relevant to the Amharic questions and retrieves passages that are `accidentally' close in the multi-lingual vector space. After we post-trained the models, the language distribution becomes more spread over different languages. For the Amharic--English model, the highest percentage of retrieved documents is in English, which is expected. For the Amharic--Arabic model, no retrieved documents are in the higher-resource language Arabic. However, we do see that the amount of retrieved documents in Amharic is larger, almost 20\%, compared to the Amharic--English model. This might indicate that the models have learned the representation of the Amharic language better but did not actually align Amharic to the high-resource language. 

After language alignment during fine-tuning, the language distribution for the Amharic--English model is still quite spread, and the language with the highest percentage is also English (46\%). When aligned with English, since it is a high-resource language, the other languages might also include texts in English, therefore the language distribution is more spread under the different languages. When we align Amharic with Arabic or Thai, the retrieved passages are mostly taken from these high-resource languages. For the Amharic--Arabic model, 98\% of the passages retrieved are Arabic. This analysis is also visualized in Figure \ref{fig:am_lang_distr} in the Appendix.

\subsection{Ablation study}
Besides our trained models and the baseline, we conducted an ablation study where we left out the TLM part during post-training and only used the MLM post-trained model; 
The MLM-trained models 
these are only post-trained on the low-resource language itself. However, the fine-tuning processes are the same, so both models are trained on question--passage alignment data. With this ablation study, we only measured the effect of having TLM in the post-training phase. 

The results are in Table \ref{ab_recall}. 
The MLM-only-trained models seem to be worse than the models trained with MLM and TLM together, but the differences are small in all settings, around 1\% point (or less).
The reason why the improvements are small might be that TLM is also used to learn the embeddings of these newly added Amharic/Khmer words with the help of the high-resource language, but these words are already learned during the MLM process. We think that the addition of TLM does not change the embeddings much, and therefore results in similar performance. Another reason might be that the language alignment is also done during the fine-tuning process, which reduces the effect of TLM and therefore results in similar performance between the models trained with or without TLM.

\begin{table}[t]
    \caption{Comparing performance of different fine-tuned models with MLM included or excluded in the post-training in the ablation study in terms of Recall@k. English is used as high-resource language in these settings. }
    \centering
    \begin{tabular}{|l||c|c|}
         \hline
         &  \textbf{R@10 (\%)} & \textbf{R@20 (\%)}  \\
         \midrule
         \multicolumn{3}{|l|}{\textbf{Amharic}} \\
         \hline
         Model$_{mlm+tlm}$ & \textbf{1.41} & \textbf{1.87} \\
         \hline
         Model$_{mlm}$ & 0.61  & 1.07  \\
         \hline
         \multicolumn{3}{|l|}{\textbf{Khmer}} \\
         \hline
         Model$_{mlm+tlm}$ & \textbf{18.64} & \textbf{22.00} \\
         \hline
         Model$_{mlm}$ & 18.42  & 21.92  \\
         \hline
    \end{tabular}
    \label{ab_recall}
\end{table}

\section{Discussion}
We found that language alignment can bring some improvements compared to the baseline, but the effect is small: the results remain low for the low-resource languages. It appears challenging for the model to learn proper representations for retrieving passages in these two languages. One problem that we identify is that Amharic and Khmer use a different writing script than their corresponding high-resource languages. This is in line with suggestions done in prior work \cite{bonab2019simulating,kumar2022mucot}. 

We found that the alignment with the English language is the best-performing one, probably because it is the language with the most data.  We think that the QA performance can be improved if we can provide more data (during post-training and fine-tuning) in these languages to train the model. Further, we found that the performance of the high-resource languages diminished due to the language alignment, probably caused by the effect known as the \textit{curse of multilinguality} \cite{pfeiffer2022lifting}. This also shows how difficult it is to improve the multilingual QA performance for low-resource languages.

The low performance of Amharic might be due to the fact that the questions and answers are taken from Wikipedia pages, and there is a chance that the Wikipedia pages in other languages do not contain the correct answers for these Amharic questions. Therefore, having passages taken from Wikipedia articles in multiple languages for retrieval, and alignment with the high-resource languages, can make the model more likely to retrieve passages in other languages than Amharic. We would expect that when more Amharic articles are retrieved, the performance would also be better. We did an additional analysis with the retrieval and found that when we only kept the Amharic passages, the performance indeed improved. However, when we have all passages in all languages, the model is more likely to retrieve passages from the high-resource languages instead of Amharic. This indicates that the model 
has learned the mapping between the low-resource and high-resource languages through question--passage alignment. 


\section{Conclusions}
We analyzed the potential of mDPR for cross-lingual answer retrieval for low-resource languages. 
We tried to improve the effectiveness of IR for two low-resource languages that do not occur in the mBERT pre-training data: Amharic and Khmer. We post-trained and fine-tuned mDPR on newly created datasets with two forms of language alignment. We show that our training approach improves the results, but we also identify additional challenges to this task: (1) the languages are not only low-resource, but also different in script type from other languages in the model; (2) the quality of the training data remains limited; (3) the evaluation approach is indirect and in-exact. This indicates that cross-lingual IR for low-resource languages is not close to being solved. We hope that the datasets we released will enable follow-up research into low-resource cross-lingual IR.

\section*{Limitations} 
\textbf{Data quality.} In our post-training alignment, we used the CCAligned dataset, and although we used translation and similarity scores above a certain threshold, there is still a chance that some of the texts in each language pair are not correctly aligned. When this is the case, the model can get confused by these not-aligned text pairs, which diminishes the performance of the model. 

\textbf{Translation quality.} Especially in the fine-tuning process, we trained the model on the translated questions, but the translation quality is not guaranteed because the languages we work with are low-resource languages, and the translation API we use is not well-trained in these languages. This limitation was acknowledged by prior work as well \cite{fei2021cross}. 

\textbf{Evaluation method.} The evaluation method we used is based on answer-level Recall. As we evaluated mDPR, which is a retrieval model, but we only have the ground truth answers for the extraction task (and no relevance labels), we can only evaluate our retrieval task with extraction labels. 
Therefore, we cannot 100\% guarantee that the percentage we show is the percentage of correct answers since we analysed that several retrieved passages are not related to the question, but the answer coincidentally appears in the passages. We can improve this in the future by providing relevance labels or transforming the task into an extraction or a generation task. 
\bibliography{custom}

\begin{thebibliography}{41}
\providecommand{\natexlab}[1]{#1}

\bibitem[{Abedissa et~al.(2023)Abedissa, Usbeck, and Assabie}]{abedissa2023amqa}
Tilahun Abedissa, Ricardo Usbeck, and Yaregal Assabie. 2023.
\newblock \href {https://arxiv.org/abs/2303.03290} {Amqa: Amharic question answering dataset}.

\bibitem[{Antony and Paul(2023)}]{antony2023question}
Betina Antony and NR~Rejin Paul. 2023.
\newblock \href {https://link.springer.com/chapter/10.1007/978-3-031-33231-9_17} {Question answering system for tamil using deep learning}.
\newblock In \emph{Speech and Language Technologies for Low-Resource Languages}, pages 244--252, Cham. Springer International Publishing.

\bibitem[{Artetxe et~al.(2020)Artetxe, Ruder, and Yogatama}]{artetxe2020cross}
Mikel Artetxe, Sebastian Ruder, and Dani Yogatama. 2020.
\newblock \href {https://doi.org/10.18653/v1/2020.acl-main.421} {On the cross-lingual transferability of monolingual representations}.
\newblock In \emph{Proceedings of the 58th Annual Meeting of the Association for Computational Linguistics}, pages 4623--4637, Online. Association for Computational Linguistics.

\bibitem[{Asai et~al.(2022)Asai, Longpre, Kasai, Lee, Zhang, Hu, Yamada, Clark, and Choi}]{asai2022mia}
Akari Asai, Shayne Longpre, Jungo Kasai, Chia-Hsuan Lee, Rui Zhang, Junjie Hu, Ikuya Yamada, Jonathan~H. Clark, and Eunsol Choi. 2022.
\newblock \href {https://doi.org/10.18653/v1/2022.mia-1.11} {{MIA} 2022 shared task: Evaluating cross-lingual open-retrieval question answering for 16 diverse languages}.
\newblock In \emph{Proceedings of the Workshop on Multilingual Information Access (MIA)}, pages 108--120, Seattle, USA. Association for Computational Linguistics.

\bibitem[{Asai et~al.(2021)Asai, Yu, Kasai, and Hajishirzi}]{asai2021one}
Akari Asai, Xinyan Yu, Jungo Kasai, and Hanna Hajishirzi. 2021.
\newblock One question answering model for many languages with cross-lingual dense passage retrieval.
\newblock In \emph{Advances in Neural Information Processing Systems}, volume~34, pages 7547--7560. Curran Associates, Inc.

\bibitem[{Banerjee et~al.(2014)Banerjee, Naskar, and Bandyopadhyay}]{banerjee2014bfqa}
Somnath Banerjee, Sudip~Kumar Naskar, and Sivaji Bandyopadhyay. 2014.
\newblock \href {https://link.springer.com/chapter/10.1007/978-3-319-10816-2_27} {Bfqa: A bengali factoid question answering system}.
\newblock In \emph{Text, Speech and Dialogue}, pages 217--224, Cham. Springer International Publishing.

\bibitem[{Bonab et~al.(2019)Bonab, Allan, and Sitaraman}]{bonab2019simulating}
Hamed Bonab, James Allan, and Ramesh Sitaraman. 2019.
\newblock Simulating clir translation resource scarcity using high-resource languages.
\newblock In \emph{Proceedings of the 2019 ACM SIGIR International Conference on Theory of Information Retrieval}, pages 129--136.

\bibitem[{Clark et~al.(2020)Clark, Choi, Collins, Garrette, Kwiatkowski, Nikolaev, and Palomaki}]{clark2020tydi}
Jonathan~H. Clark, Eunsol Choi, Michael Collins, Dan Garrette, Tom Kwiatkowski, Vitaly Nikolaev, and Jennimaria Palomaki. 2020.
\newblock \href {https://doi.org/10.1162/tacl_a_00317} {{T}y{D}i {QA}: A benchmark for information-seeking question answering in typologically diverse languages}.
\newblock \emph{Transactions of the Association for Computational Linguistics}, 8:454--470.

\bibitem[{Conneau and Lample(2019)}]{conneau2019cross}
Alexis Conneau and Guillaume Lample. 2019.
\newblock \href {https://proceedings.neurips.cc/paper_files/paper/2019/file/c04c19c2c2474dbf5f7ac4372c5b9af1-Paper.pdf} {Cross-lingual language model pretraining}.
\newblock In \emph{Advances in Neural Information Processing Systems}, volume~32. Curran Associates, Inc.

\bibitem[{Darvishi et~al.(2023)Darvishi, Shahbodaghkhan, Abbasiantaeb, and Momtazi}]{darvishi2023pquad}
Kasra Darvishi, Newsha Shahbodaghkhan, Zahra Abbasiantaeb, and Saeedeh Momtazi. 2023.
\newblock \href {https://doi.org/10.1016/j.csl.2023.101486} {Pquad: A persian question answering dataset}.
\newblock \emph{Computer Speech \& Language}, 80:101486.

\bibitem[{Das et~al.(2022)Das, Mandal, Danial, Pal, and Saha}]{das2022improvement}
Arijit Das, Jaydeep Mandal, Zargham Danial, Alok~Ranjan Pal, and Diganta Saha. 2022.
\newblock \href {https://link.springer.com/article/10.1007/s12046-021-01765-3} {An improvement of bengali factoid question answering system using unsupervised statistical methods}.
\newblock \emph{S{\=a}dhan{\=a}}, 47(1):2.

\bibitem[{Das and Saha(2022)}]{das2022question}
Arijit Das and Diganta Saha. 2022.
\newblock \href {https://onlinelibrary.wiley.com/doi/abs/10.1002/9781119857686.ch10} {Question answering system using deep learning in the low resource language bengali}.
\newblock \emph{Convergence of Deep Learning In Cyber-IoT Systems and Security}, pages 207--230.

\bibitem[{Devlin et~al.(2019)Devlin, Chang, Lee, and Toutanova}]{devlin2019bert}
Jacob Devlin, Ming-Wei Chang, Kenton Lee, and Kristina Toutanova. 2019.
\newblock \href {https://doi.org/10.18653/v1/N19-1423} {{BERT}: Pre-training of deep bidirectional transformers for language understanding}.
\newblock In \emph{Proceedings of the 2019 Conference of the North {A}merican Chapter of the Association for Computational Linguistics: Human Language Technologies, Volume 1 (Long and Short Papers)}, pages 4171--4186, Minneapolis, Minnesota. Association for Computational Linguistics.

\bibitem[{Do et~al.(2021)Do, Phan, and Gupta}]{do2021developing}
Phuc Do, Truong H.~V. Phan, and Brij~B. Gupta. 2021.
\newblock \href {https://doi.org/10.1145/3453651} {{Developing a Vietnamese Tourism Question Answering System Using Knowledge Graph and Deep Learning}}.
\newblock \emph{ACM Trans. Asian Low-Resour. Lang. Inf. Process.}, 20(5).

\bibitem[{El-Kishky et~al.(2020)El-Kishky, Chaudhary, Guzm{\'a}n, and Koehn}]{el2020ccaligned}
Ahmed El-Kishky, Vishrav Chaudhary, Francisco Guzm{\'a}n, and Philipp Koehn. 2020.
\newblock \href {https://doi.org/10.18653/v1/2020.emnlp-main.480} {{CCA}ligned: A massive collection of cross-lingual web-document pairs}.
\newblock In \emph{Proceedings of the 2020 Conference on Empirical Methods in Natural Language Processing (EMNLP)}, pages 5960--5969, Online. Association for Computational Linguistics.

\bibitem[{Fei et~al.(2021)Fei, Yu, and Li}]{fei2021cross}
Hongliang Fei, Tan Yu, and Ping Li. 2021.
\newblock Cross-lingual cross-modal pretraining for multimodal retrieval.
\newblock In \emph{Proceedings of the 2021 Conference of the North American Chapter of the Association for Computational Linguistics: Human Language Technologies}, pages 3644--3650.

\bibitem[{Huang et~al.(2023)Huang, Yu, and Allan}]{huang2023improving}
Zhiqi Huang, Puxuan Yu, and James Allan. 2023.
\newblock Improving cross-lingual information retrieval on low-resource languages via optimal transport distillation.
\newblock In \emph{Proceedings of the Sixteenth ACM International Conference on Web Search and Data Mining}, pages 1048--1056.

\bibitem[{Karpukhin et~al.(2020)Karpukhin, Oguz, Min, Lewis, Wu, Edunov, Chen, and Yih}]{karpukhin2020dense}
Vladimir Karpukhin, Barlas Oguz, Sewon Min, Patrick Lewis, Ledell Wu, Sergey Edunov, Danqi Chen, and Wen-tau Yih. 2020.
\newblock \href {https://doi.org/10.18653/v1/2020.emnlp-main.550} {Dense passage retrieval for open-domain question answering}.
\newblock In \emph{Proceedings of the 2020 Conference on Empirical Methods in Natural Language Processing (EMNLP)}, pages 6769--6781, Online. Association for Computational Linguistics.

\bibitem[{Kumar et~al.(2022)Kumar, Gehlot, Mullappilly, and Nandakumar}]{kumar2022mucot}
Gokul~Karthik Kumar, Abhishek Gehlot, Sahal~Shaji Mullappilly, and Karthik Nandakumar. 2022.
\newblock \href {https://doi.org/10.18653/v1/2022.dravidianlangtech-1.3} {{M}u{C}o{T}: Multilingual contrastive training for question-answering in low-resource languages}.
\newblock In \emph{Proceedings of the Second Workshop on Speech and Language Technologies for Dravidian Languages}, pages 15--24, Dublin, Ireland. Association for Computational Linguistics.

\bibitem[{Kwiatkowski et~al.(2019)Kwiatkowski, Palomaki, Redfield, Collins, Parikh, Alberti, Epstein, Polosukhin, Devlin, Lee, Toutanova, Jones, Kelcey, Chang, Dai, Uszkoreit, Le, and Petrov}]{kwiatkowski2019natural}
Tom Kwiatkowski, Jennimaria Palomaki, Olivia Redfield, Michael Collins, Ankur Parikh, Chris Alberti, Danielle Epstein, Illia Polosukhin, Jacob Devlin, Kenton Lee, Kristina Toutanova, Llion Jones, Matthew Kelcey, Ming-Wei Chang, Andrew~M. Dai, Jakob Uszkoreit, Quoc Le, and Slav Petrov. 2019.
\newblock \href {https://doi.org/10.1162/tacl_a_00276} {Natural questions: A benchmark for question answering research}.
\newblock \emph{Transactions of the Association for Computational Linguistics}, 7:452--466.

\bibitem[{Le et~al.(2022)Le, Nguyen, Le~Thanh, and Nguyen}]{le2022vimqa}
Khang Le, Hien Nguyen, Tung Le~Thanh, and Minh Nguyen. 2022.
\newblock \href {https://aclanthology.org/2022.lrec-1.700} {{VIMQA}: A {V}ietnamese dataset for advanced reasoning and explainable multi-hop question answering}.
\newblock In \emph{Proceedings of the Thirteenth Language Resources and Evaluation Conference}, pages 6521--6529, Marseille, France. European Language Resources Association.

\bibitem[{Longpre et~al.(2021)Longpre, Lu, and Daiber}]{longpre2021mkqa}
Shayne Longpre, Yi~Lu, and Joachim Daiber. 2021.
\newblock \href {https://doi.org/10.1162/tacl_a_00433} {Mkqa: A linguistically diverse benchmark for multilingual open domain question answering}.
\newblock \emph{Transactions of the Association for Computational Linguistics}, 9:1389--1406.

\bibitem[{Mozafari et~al.(2022)Mozafari, Kazemi, Moradi, Nematbakhsh, and Jafari}]{mozafari2022peransel}
Jamshid Mozafari, Arefeh Kazemi, Parham Moradi, Mohammad~Ali Nematbakhsh, and Sajad Jafari. 2022.
\newblock \href {https://doi.org/10.1155/2022/3661286} {Peransel: A novel deep neural network-based system for persian question answering}.
\newblock \emph{Intell. Neuroscience}, 2022.

\bibitem[{Ogundepo et~al.(2022)Ogundepo, Zhang, Sun, Duh, and Lin}]{ogundepo2022africlirmatrix}
Odunayo Ogundepo, Xinyu Zhang, Shuo Sun, Kevin Duh, and Jimmy Lin. 2022.
\newblock Africlirmatrix: Enabling cross-lingual information retrieval for african languages.
\newblock In \emph{Proceedings of the 2022 Conference on Empirical Methods in Natural Language Processing}, pages 8721--8728.

\bibitem[{Pan et~al.(2021)Pan, Hang, Qi, Shah, Potdar, and Yu}]{pan2021multilingual}
Lin Pan, Chung-Wei Hang, Haode Qi, Abhishek Shah, Saloni Potdar, and Mo~Yu. 2021.
\newblock \href {https://doi.org/10.18653/v1/2021.naacl-main.20} {Multilingual {BERT} post-pretraining alignment}.
\newblock In \emph{Proceedings of the 2021 Conference of the North American Chapter of the Association for Computational Linguistics: Human Language Technologies}, pages 210--219, Online. Association for Computational Linguistics.

\bibitem[{Pfeiffer et~al.(2022)Pfeiffer, Goyal, Lin, Li, Cross, Riedel, and Artetxe}]{pfeiffer2022lifting}
Jonas Pfeiffer, Naman Goyal, Xi~Lin, Xian Li, James Cross, Sebastian Riedel, and Mikel Artetxe. 2022.
\newblock \href {https://doi.org/10.18653/v1/2022.naacl-main.255} {Lifting the curse of multilinguality by pre-training modular transformers}.
\newblock In \emph{Proceedings of the 2022 Conference of the North American Chapter of the Association for Computational Linguistics: Human Language Technologies}, pages 3479--3495, Seattle, United States. Association for Computational Linguistics.

\bibitem[{Phan and Do(2021)}]{phan2021building}
Trung Phan and Phuc Do. 2021.
\newblock \href {https://doi.org/10.1007/s00521-021-06126-z} {{Building a Vietnamese Question Answering System Based on Knowledge Graph and Distributed CNN}}.
\newblock \emph{Neural Comput. Appl.}, 33(21):14887–14907.

\bibitem[{Pramana and Prasojo(2022)}]{pramana2022improving}
Ryan Pramana and Radityo~Eko Prasojo. 2022.
\newblock \href {https://doi.org/10.1109/ICoICT55009.2022.9914847} {Improving low-resource question answering with cross-lingual data augmentation strategies}.
\newblock In \emph{2022 10th International Conference on Information and Communication Technology (ICoICT)}, pages 110--115. IEEE.

\bibitem[{Ravva et~al.(2020)Ravva, Urlana, and Shrivastava}]{ravva2020avadhan}
Priyanka Ravva, Ashok Urlana, and Manish Shrivastava. 2020.
\newblock \href {https://doi.org/10.1145/3371158.3371193} {Avadhan: System for open-domain telugu question answering}.
\newblock In \emph{Proceedings of the 7th ACM IKDD CoDS and 25th COMAD}, CoDS COMAD 2020, page 234–238, New York, NY, USA. Association for Computing Machinery.

\bibitem[{Riabi et~al.(2021)Riabi, Scialom, Keraron, Sagot, Seddah, and Staiano}]{riabi2021synthetic}
Arij Riabi, Thomas Scialom, Rachel Keraron, Beno{\^\i}t Sagot, Djam{\'e} Seddah, and Jacopo Staiano. 2021.
\newblock \href {https://doi.org/10.18653/v1/2021.emnlp-main.562} {Synthetic data augmentation for zero-shot cross-lingual question answering}.
\newblock In \emph{Proceedings of the 2021 Conference on Empirical Methods in Natural Language Processing}, pages 7016--7030, Online and Punta Cana, Dominican Republic. Association for Computational Linguistics.

\bibitem[{Singh et~al.(2019)Singh, McCann, Keskar, Xiong, and Socher}]{singh2019xlda}
Jasdeep Singh, Bryan McCann, Nitish~Shirish Keskar, Caiming Xiong, and Richard Socher. 2019.
\newblock \href {https://arxiv.org/abs/1905.11471} {Xlda: Cross-lingual data augmentation for natural language inference and question answering}.

\bibitem[{Thompson(2023)}]{Michael2023Thai}
Michael Thompson. 2023.
\newblock \href {https://ling-app.com/th/thai-vs-cambodian/} {Thai vs. cambodian: How similar are these 2 incredible languages? - ling app}.

\bibitem[{Toraman et~al.(2023)Toraman, Yilmaz, {\c{S}}ahinu{\c{c}}, and Ozcelik}]{toraman2023impact}
Cagri Toraman, Eyup~Halit Yilmaz, Furkan {\c{S}}ahinu{\c{c}}, and Oguzhan Ozcelik. 2023.
\newblock \href {https://doi.org/10.1145/3578707} {Impact of tokenization on language models: An analysis for turkish}.
\newblock \emph{ACM Transactions on Asian and Low-Resource Language Information Processing}, 22(4):1--21.

\bibitem[{Vemula et~al.(2022)Vemula, Nuthi, and Srivastava}]{vemula2022tequad}
Rakesh Vemula, Mani Nuthi, and Manish Srivastava. 2022.
\newblock \href {https://aclanthology.org/2022.icon-main.36} {{T}e{Q}u{AD}:{T}elugu question answering dataset}.
\newblock In \emph{Proceedings of the 19th International Conference on Natural Language Processing (ICON)}, pages 300--307, New Delhi, India. Association for Computational Linguistics.

\bibitem[{Viriyayudhakorn and Polpanumas(2021)}]{Viriyayudhakorn2021iapp}
Kobkrit Viriyayudhakorn and Charin Polpanumas. 2021.
\newblock \href {https://doi.org/10.5281/zenodo.4539916} {iapp\_wiki\_qa\_squad}.

\bibitem[{Vuong et~al.(2023)Vuong, Nguyen, Nguyen, Nguyen, and Phan}]{vuong2023improving}
Thi-Hai-Yen Vuong, Ha-Thanh Nguyen, Quang-Huy Nguyen, Le-Minh Nguyen, and Xuan-Hieu Phan. 2023.
\newblock \href {https://arxiv.org/abs/2306.04841} {{Improving Vietnamese Legal Question--Answering System based on Automatic Data Enrichment}}.

\bibitem[{Wang et~al.(2020)Wang, Mayhew, Roth et~al.}]{wang2020extending}
Zihan Wang, Stephen Mayhew, Dan Roth, et~al. 2020.
\newblock \href {https://doi.org/10.48550/arXiv.2004.13640} {Extending multilingual bert to low-resource languages}.
\newblock \emph{arXiv preprint arXiv:2004.13640}.

\bibitem[{Wanjawa et~al.(2023)Wanjawa, Wanzare, Indede, Mconyango, Muchemi, and Ombui}]{wanjawa2023kenswquad}
Barack~W. Wanjawa, Lilian D.~A. Wanzare, Florence Indede, Owen Mconyango, Lawrence Muchemi, and Edward Ombui. 2023.
\newblock \href {https://doi.org/10.1145/3578553} {Kenswquad—a question answering dataset for swahili low-resource language}.
\newblock \emph{ACM Trans. Asian Low-Resour. Lang. Inf. Process.}, 22(4).

\bibitem[{Wu et~al.(2022)Wu, Zhu, Zhang, Zhuang, and Feng}]{wu2022enhancing}
Linjuan Wu, Jiazheng Zhu, Xiaowang Zhang, Zhiqiang Zhuang, and ZhiYong Feng. 2022.
\newblock \href {https://doi.org/10.1007/978-3-031-11217-1_13} {Enhancing low-resource languages question answering with syntactic graph}.
\newblock In \emph{Database Systems for Advanced Applications. DASFAA 2022 International Workshops}, pages 175--188, Cham. Springer International Publishing.

\bibitem[{Yarmohammadi et~al.(2019)Yarmohammadi, Ma, Hisamoto, Rahman, Wang, Xu, Povey, Koehn, and Duh}]{yarmohammadi2019robust}
Mahsa Yarmohammadi, Xutai Ma, Sorami Hisamoto, Muhammad Rahman, Yiming Wang, Hainan Xu, Daniel Povey, Philipp Koehn, and Kevin Duh. 2019.
\newblock Robust document representations for cross-lingual information retrieval in low-resource settings.
\newblock In \emph{Proceedings of Machine Translation Summit XVII: Research Track}, pages 12--20.

\bibitem[{Yu et~al.(2021)Yu, Fei, and Li}]{yu2021cross}
Puxuan Yu, Hongliang Fei, and Ping Li. 2021.
\newblock Cross-lingual language model pretraining for retrieval.
\newblock In \emph{Proceedings of the Web Conference 2021}, pages 1029--1039.

\end{thebibliography}

\newpage
\appendix

\section{Appendix}
\label{sec:appendix}

\subsection{Thai datasets}

We combine two datasets into a Thai dataset with 8,432 examples. Statistics are shown in Table~\ref{tab:thai_data}.

\begin{table}[h]
    \centering
        \caption{Dataset Distribution for Thai and Final Combined Data}
    \begin{tabular}{lr}
        \toprule
        Dataset & \# Examples \\
        \midrule
        Thai total & 8,432  \\
        Thai dev (8\%) & 675 \\
        Thai train & 7,757 \\
        \midrule
        Final training data with Thai & 120,880 \\
         Final development data with Thai & 5,203 \\
        \bottomrule
    \end{tabular}
    \label{tab:thai_data}
\end{table}

\subsection{Finetuning and data processing details}\label{sec:details}

We use the mDPR training script\footnote{\url{https://github.com/mia-workshop/MIA-Shared-Task-2022/tree/main/baseline/mDPR}\label{mdpr-script}} provided in the MIA repository with the training parameter values 
of \textbf{replication 1} (which uses the hyperparameter values of DPR \cite{karpukhin2020dense}). 
Due to limited GPU resources, we use a single GPU to train the models and apply a half-precision floating point format (FP16) instead of a single-precision floating point format (FP32) during training.
\footnote{\url{https://docs.nvidia.com/deeplearning/performance/mixed-precision-training/index.html}} 
We store the last trained checkpoints as our final models. 

After training, we use our trained mDPR models to generate dense embeddings for the Wikipedia context passages. The Wikipedia context passages consist of the processed 100-token length passages\footnote{\url{https://nlp.cs.washington.edu/xorqa/cora/models/mia2022_shared_task_all_langs_w100.tsv}\label{w100}} as provided in the MIA repository. However, the passages provided in the MIA repository do not contain passages in Thai and Amharic, therefore, we add the passages in Thai and Amharic to the collection. The Thai passages are obtained from the Wikipedia context passages\footnote{\url{https://nlp.cs.washington.edu/xorqa/cora/models/all_w100.tsv}} provided in the CORA repository.\footnote{\url{https://github.com/AkariAsai/CORA/tree/main}\label{cora}} The Amharic passages are obtained from the Wikipedia dumps in Amharic\footnote{\url{https://archive.org/download/amwiki-20210801}} (processed as described in the MIA repository).

\subsection{Tokenization details}

Since Khmer does not use whitespace between words, the mBERT tokenizer was not able to split the Khmer sentence into tokens. This results in one unknown token for the Khmer sentence part that is not separated by whitespace. When we separate the Khmer sentence by whitespace (using khmer-nltk) and then feed the separated sentence into the kmBERT tokenizer, it successfully tokenizes the Khmer sentence. 

We noticed that our extended mBERT tokenizer, initialized with AutoTokenizer using the most recent version of Transformers, is able to tokenize the Khmer sentence without it being separated by whitespace. However, during finetuning, we use the same Transformers’ version as given in the MIA repository (v3.0.2). We therefore keep the TLM alignment the same as the question--passage alignment, by first tokenizing the Khmer sentence using khmer-nltk. For MLM post-training, we did use the most recent version (v4.33.3 at the time) of Transformers, so we were able to use the kmBERT tokenizer directly to tokenize the text, and the resulting tokens are similar to that of khmer-nltk.

\begin{table}[h]
    \centering
        \caption{Document Distribution for Amharic and Khmer}
    \begin{tabular}{lrrrr}
        \toprule
        Language & Total Docs & Training & Validation & Test \\
        \midrule
        Amharic & 46,066 & 43,762 & 921 & 1,383 \\
        Khmer   & 71,994 & 68,394 & 1,439 & 2,161 \\
        \bottomrule
    \end{tabular}
    \label{tab:doc_distribution}
\end{table}

\onecolumn

\begin{table*}[h]
\caption{Tokenization examples. }
\label{tab:tokenization}
\begin{tabular}{p{2.5cm}|p{13.5cm}}
\toprule
    \multicolumn{2}{l}{\textbf{Amharic}}   \\
    \midrule
    Original & \includegraphics[width=1.3\columnwidth]{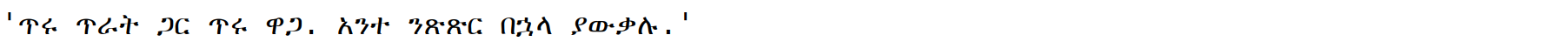} \\ 
    \midrule
    Tokenized w/ mBERT &  {\small\texttt{['[UNK]', '[UNK]', '[UNK]', '[UNK]', '[UNK]', '.', '[UNK]', '[UNK]', '[UNK]', '[UNK]', '.']}}\\ 
    \midrule
    Tokenized w/ amBERT & \includegraphics[width=\columnwidth]{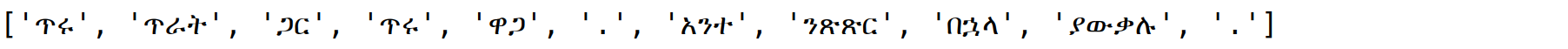} \\
\toprule
    \multicolumn{2}{l}{\textbf{Khmer}}  \\
        \midrule
    Original & \includegraphics[width=1.3\columnwidth]{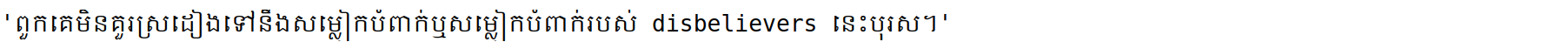} \\
\midrule
Tokenized w/ mBERT & {\small\texttt{['[UNK]', 'dis', '\#\#beli', '\#\#ever', '\#\#s', '[UNK]', '[UNK]']}} \\
\midrule
    Tokenized w/ kmBERT & \includegraphics[width=0.9\columnwidth]{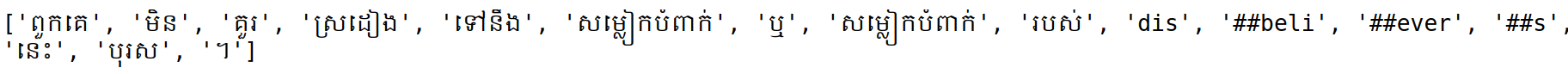}\\
    \bottomrule
\end{tabular}
\end{table*}

\begin{table*}[h]
\caption{Evaluation of the mDPR+mGEN results for the different models on the XOR-TyDi development data (Thai extended).}
    \centering
    \begin{tabular}{|l||c|c|c|c|c|c|c|c||c|c|c|}
        \hline
                        & \multicolumn{8}{|c||}{\textbf{XOR-TyDi Language F1}} & \multicolumn{3}{|c|}{\textbf{Average}}\\
                        & ar    & bn    & fi    & ja    & ko    & ru    & te    & th    & \textbf{F1} & \textbf{EM} & \textbf{BLEU}\\
        \hline \hline
        Baseline        & 48.43 & 26.36 & 41.91 & 37.23 & 26.71 & 34.48 & 39.04 & 28.28 & 35.30 & 26.51 & 25.35 \\
        \hline 
        \multicolumn{12}{|l|}{\textbf{Amharic}} \\
        \hline
        Model$_{am-en}$ & 43.53 & 22.66 & 35.43 & 34.62 & 26.05 & 30.82 & 35.62 & 25.64 & 31.80 & 23.78 & 22.45 \\
        \hline
        Model$_{am-ar}$ & 43.86 & 23.00 & 38.80 & 33.68 & 23.04 & 29.55 & 35.64 & 25.90 & 31.68 & 23.81 & 22.38 \\
        \hline
        Model$_{am-th}$ & 42.62 & 23.49 & 37.21 & 35.32 & 21.43 & 32.59 & 36.91 & 28.60 & 32.27 & 24.23 & 23.20 \\
        \hline 
        \multicolumn{12}{|l|}{\textbf{Khmer}} \\
        \hline
        Model$_{km-en}$ & 43.62 & 22.12 & 37.81 & 33.34 & 23.20 & 30.85 & 37.46 & 25.06 & 31.68 & 23.75 & 22.60 \\
        \hline
        Model$_{km-ar}$ & 43.15 & 18.58 & 36.65 & 34.53 & 20.70 & 30.88 & 33.12 & 25.70 & 30.41 & 22.58 & 21.75 \\
        \hline
        Model$_{km-th}$ & 44.21 & 22.88 & 36.27 & 33.02 & 23.70 & 32.44 & 36.05 & 29.01 & 32.20 & 24.04 & 23.13 \\
        \hline
    \end{tabular}
    \label{xor_res}
\end{table*}

\begin{table*}[h]
    \caption{Retrieval performance of different mDPR models for Amharic and Khmer in terms of ROUGE-1.
    }
    \centering
    \begin{tabular}{|l||cccc|cccc|}
         \toprule
         &  \multicolumn{4}{|c|}{\textbf{Top10 ROUGE-1 (\%)}} & \multicolumn{4}{|c|}{\textbf{Top20 ROUGE-1 (\%)}}  \\
\toprule
\textbf{Amharic} &  & am-en & am-ar & am-th &  & am-en & am-ar & am-th\\
\hline
mBERT & 7.29 &  &  &  & 7.84 &  &  & \\
+ Posttraining &  & 10.17 & \textbf{11.57} & 10.69 &  & 11.61 & \textbf{14.41} & 12.58\\
+ FT align  &  & 8.94 & 7.80 & 7.25 &  & 9.78 & 8.39 & 7.79\\
\midrule
\textbf{Khmer}  &  & km-en & km-ar & km-th &  & km-en & km-ar & km-th\\
\hline
mBERT & 15.17 &  &  &  & 19.34 &  &  & \\
+ Posttraining &  & 13.83 & 12.51 & 16.81 &  & 17.25 & 15.89 & 20.93\\
+ FT align  &  & \textbf{18.79} & 15.00 & 15.22 &  & \textbf{23.05} & 18.63 & 19.39\\
\bottomrule
    \end{tabular}
    \label{rouge_results}
\end{table*}

\begin{figure*}[t]
    \centering
    \includegraphics[width=0.4\linewidth]{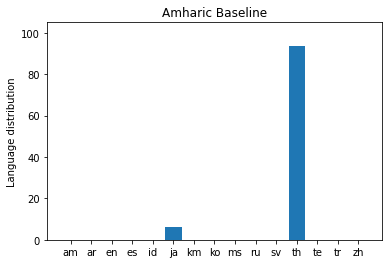}
    \includegraphics[width=0.4\linewidth]{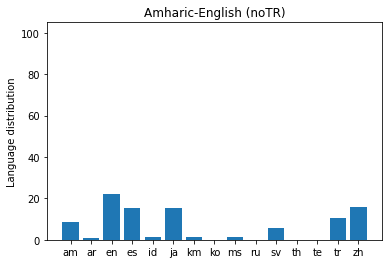}
    \includegraphics[width=0.4\linewidth]{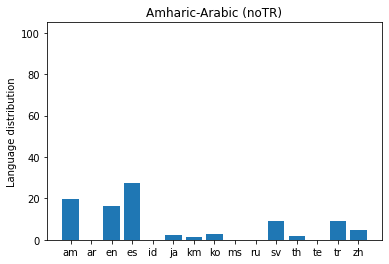}
    \includegraphics[width=0.4\linewidth]{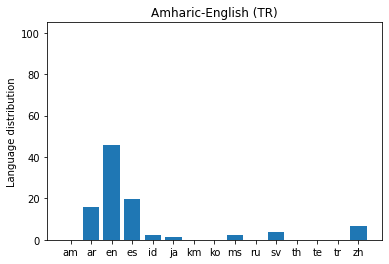}
    \includegraphics[width=0.4\linewidth]{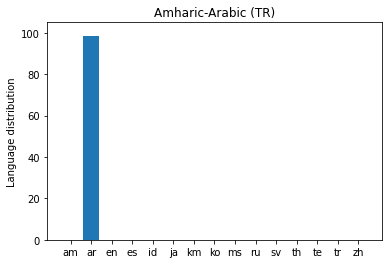}
    \caption{Language distribution of the top 20 retrieved documents for the different Amharic models. NoTR is the setting where the model is not trained on the translated training data for the low-resource language during fine-tuning. TR is the setting where the model is trained on the translated training data for the low-resource language during fine-tuning. We only report languages where at least 1\% of the retrieved documents appear in that language.} 
    \label{fig:am_lang_distr}
\end{figure*}

\subsection{Qualitative analysis}
We looked into the top 10 retrieved passages in the TL\_EN setting for each language pair to see to what extent the retrieved passages are related to the questions. Here, we randomly took 20 questions and the top 10 passages retrieved for those questions where at least an answer is found in the passages for that question. For the Amharic baseline, we found that the retrieved passages are almost the same for each question, and not specifically related to the questions. For the Amharic--English model, in 8 out of 20 questions, the passages are (somewhat) related to the questions. For example, for the question ``How many times did athlete Haile win an Olympic gold medal?'', the Amharic--English model retrieves passages related to the Olympics. 

For the Amharic--Arabic model, in 9 out of 20 questions, we have (somewhat) related passages. For the other questions, most passages share the same topic but are not related to the questions. This might be because the dense embeddings of these topics might be close to the dense embeddings of the questions, which confused the model for retrieving the right passages. When qualitatively looking at the relatedness between the passages and the questions based on this small sample, we judge the Amharic--Arabic model as comparable to the Amharic--English model.

\end{document}